# Evaluation of Angular Dispersion for Various Propagation Environments in Emerging 5G Systems

Jan M. Kelner, Cezary Ziółkowski, and Bogdan Uljasz
Institute of Telecommunications, Faculty of Electronics, Military University of Technology,
Warsaw, Poland
{jan.kelner, cezary.ziolkowski, bogdan.uljasz}@wat.edu.pl

*Abstract*—Angular dispersion is the effect of a multi-path propagation observed in received signals. An assessment of this phenomenon is particularly important from the viewpoint of emerging fifth generation (5G) communication systems. In these systems, using the beam-forming and massive multiple-input multiple-output antenna arrays are planned. This phenomenon also has a negative impact on direction finding and older generation communication systems used in an urban environment. In this paper, we present the angular dispersion evaluation for various propagation environments based on simulation studies. This analysis is carried out for different environment types defined in the 3GPP standard model for a selected frequency. In this case, the angular spread is determinated based on the power angular spectrum. This parameter is the basis for the influence evaluation of the propagation environment on the received signal angular dispersion.

## I. Introduction

A development direction of emerging wireless systems is to provide a greater system capacity and transmission bitrate. Majority of a current third (3G) and fourth generation (4G) mobile systems [1] and wireless WiFi networks in the IEEE 802.11 b/g standards [2] use ultra high frequency (UHF) bands. Limited spectral resources in the UHF range are the reason for the use of higher frequency ranges, including super (SHF) and extremely high frequency (EHF). In these bands, WiFi in the IEEE 802.11 n/ac/ad standards [2], fifth generation (5G) communication systems [3] and Internet of things [4] work or will operate in the near future.

However, the use of higher frequency ranges, especially millimeter waves (EHF), has its drawbacks. The main difficulty associated with their use is a large path loss and small range of a transmitter (Tx) [5]. Hence, radio networks operating in these bands have a higher density of nodes/cells per area unit than networks operating in the UHF. In addition, tremendously high frequency (THF) and infrared optic ranges are increasingly used for line-of-sight (LOS) conditions. These are so-called terahertz [6][7] and free space optics (FSO) [8][9] communication systems.

A common feature of the EHF and THF ranges is the use of geometric optics principles [10] in propagation phenomena modeling. The wave propagation modeling for the THF and optical frequency ranges is relatively simple, as it mainly concerns LOS conditions. For the EHF, the millimeter wave propagation can also occur under non-LOS (NLOS) conditions. In this case, the additional problem is a multipath propagation phenomenon. This is associated with dispersions in time and angle domains that can be observe in a received signal. For modeling these phenomena, geometry-based propagation models are used, e.g., [11]. Geometric optics is the basis of these models.

The evaluation of the angular dispersion of the received signals consists in determining a distribution of angle of arrival angle (AOA) at surroundings of a receiver (Rx). In this case, the analysis reduces to the evaluation of propagation path trajectories in the presence of scatterers. Appropriate geometric structures are used to map scatterer positions on the plane (2D) or in space (3D). Shapes of scattering areas, their location in relation to the Tx and Rx positions, and a density distribution of the scatterers are the criteria that differentiate individual models.

Currently, 3D models are more popular, e.g., [11]. They provide to evaluate the angular dispersion in the azimuth and elevation planes. However, results presented in a literature, e.g., [12][13], show unequivocally that the phenomenon of the angular dispersion is more visible in the azimuth plane. In the elevation plane, a parameter defining this dispersion, i.e., rms angle spread (AS), is usually equal to a few degrees. For this reason, the assessment of the angular dispersion presented in this paper focuses only on the azimuth plane.

The previous wireless systems were based mainly on omnidirectional or sectorial antenna systems. Diversification of radio resources also in the field of space has forced the use of spatial multiplexing techniques such as multiple-input multiple-output (MIMO) [14]. In the emerging 5G systems, more complex antenna techniques are planned to use [3], e.g., wideband beamforming [15], massive-MIMO [16], active phased array antenna (APAA), and massive APAA [17]. In practical terms, a single beam can be modeled as a narrow-beam directional antenna. In the future communication systems, in addition to antenna arrays, the millimeter waves also enforces the use of singular directional antennas, which are characterized by low half power beamwidths (HPBWs) and high gains [5].

The purpose of this paper is to evaluate the angular dispersion for different environment types. This analysis is carried out for 38-39 GHz, which is planned to use in the upcoming 5G systems. In the assessment based on simulation studies, we use the multi-elliptical propagation model (MPM) [18]. The choice of this model results from two premises. Firstly, MPM is characterized by the best approximation of measurement data available in a literature [19]. Secondly, MDM is one of few models that considers the transmitting and receiving antenna patterns. As mentioned above, considering the antenna patterns in the analysis is very important in modeling the emerging 5G communication systems.

The remainder of this paper is organized as follows. Section II describes the method of modeling the angular dispersion. Assumptions for simulation studies and AS assessment for various propagation environments are presented in Sections III and IV, respectively. In section V, the summary is shown.

## II. ANGULAR DISPERSION MODELING

The basis of MPM is the multi-elliptical structure, which defines potential locations of the scatterers. In this case, the Tx and Rx are located in the ellipse foci. This approach was first proposed by Parsons and Bajwa [20]. A cluster structure of empirical power delay profiles (PDPs) or spectrum (PDSs) is the premise for this. Therefore, the basic input data for MPM is PDP/PDS. The dimensions of the confocal ellipses results from the delays of the analyzed PDP. The MPM geometry is shown in Fig. 1.

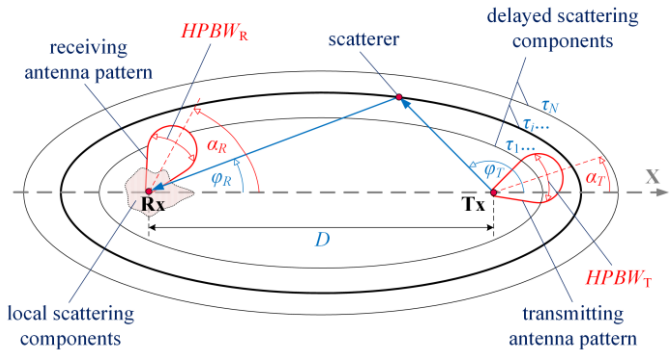

Fig. 1. Geometry of MPM.

Delayed scattering components associated with the multi-elliptical structure are the core of this geometry. In MPM, we assume additionally the occurrence possibility of a direct path and local scattering around the receiving antenna. In this case, the von Mises distribution [21] is used for the local scattering components. However, for LOS conditions, the Rice factor determines the power division between the direct path component and local scattering.

In [19], MPM for omnidirectional antennas is described. Consideration of the transmitting and receiving antenna patterns was first presented for the 3D model, called the multi-ellipsoid model, in [22] and [23], respectively. Its simplification to the azimuth plane shown in [24][18] is the basis for evaluating the angular dispersion in this paper.

The angular dispersion can be analyzed in surroundings of the Rx or at the output of the receiving antenna. In the first case, we are talking about signal analysis at the so-called reception point that is independent of the receiving antenna pattern, but considers the influence of the transmitting antenna and environment. In practice, we can assume that the receiving antenna is omnidirectional and isotropic. In the second case, the analysis concerns the signal at the output of the receiving antenna or at the Rx input. Then, the signal is already changed by the receiving antenna pattern.

Analysis of this phenomena in a literature shows that directional receiving antennas significantly reduce the angular dispersion. Therefore, the analysis of this phenomenon for various propagation environments should be carried out at the reception point, not at the directional Rx antenna output. However, comparison of ASs for both cases is presented.

## III. SIMULATION ASSUMPTIONS

The PDP is basic input data for MPM. This channel transmission characteristic and rms delay spread (DS) describe a dispersion in time domain. In many standard models, e.g., [13][25], DS is used to classify the propagation environment types. In the assessment, we use PDPs and DSs defined by the 3GPP standard for the frequency range, 0.5-100 GHz [13].

In simulation studies, we use two normalized PDPs defined as tapped delay lines (TDLs) for NLOS conditions, i.e., TDL-A [13, Table 7.7.2-1] and TDL-B [13, Table 7.7.2-2]. Based on [13, Table 7.7.3-2], we adapt DSs for 39 GHz, which define different environments: indoor office, urban micro (UMi) street-canyon, urban macro (UMa), and UMi / UMa outdoor-to-indoor (O2I). These DSs are shown in Table I.

TABLE I. DSs FOR DIFFERENT PROPAGATION SCENARIOS FOR 39 GHz

| Environment Type | PDP Type | DS (ns) | Scenario |
|---|---|---|---|
| **Indoor office** | Short-delay profile | 16 | Sc1 |
| | Normal-delay profile | 18 | Sc2 |
| | Long-delay profile | 41 | Sc3 |
| **UMi street-canyon** | Short-delay profile | 30 | Sc4 |
| | Normal-delay profile | 61 | Sc5 |
| | Long-delay profile | 297 | Sc6 |
| **UMa** | Short-delay profile | 78 | Sc7 |
| | Normal-delay profile | 249 | Sc8 |
| | Long-delay profile | 786 | Sc9 |
| **UMi / UMa O2I** | Normal-delay profile | 240 | Sc10 |
| | Long-delay profile | 616 | Sc11 |

Based on [5], the following parameters are used: $HPBW_A = 7.8°$, $G_A = 25$ dBi, and $HPBW_B = 49.4°$, $G_B = 13.3$ dBi for narrow-beam (NBA) and wide-beam (WBA) antennas, respectively.

Due to the varied propagation environment, we assume various Tx-Rx distances: 50, 100, 200, and 100 m for indoor office, UMi street-canyon, UMa, and UMi / UMa O2I, respectively.

## IV. ASSESSMENT OF ANGLE SPREAD FOR DIFFERENT PROPAGATION ENVIRONMENTS

The evaluation of the angular dispersion for various environment types is based on AS, $\sigma_\varphi$. This measure is defined as [26]

$$\sigma_\varphi = \sqrt{\int_{-180°}^{180°} \varphi_R^2 f(\varphi_R) d\varphi_R - \left(\int_{-180°}^{180°} \varphi_R f(\varphi_R) d\varphi_R\right)^2} \quad (1)$$

where $\varphi_R$ is AOA and $f(\varphi_R)$ is the distribution of AOA.

The estimation methods of the AOA distribution for the reception point and output of the receiving antenna are described in [24] and [18], respectively.

The AS changes at the reception point versus the direction, $\alpha_T$, of the selective transmitting antenna are shown in Figs. 2-9 for the analyzed environment types and two types of antennas, respectively. In all cases, we assume $\alpha_R = 0$.

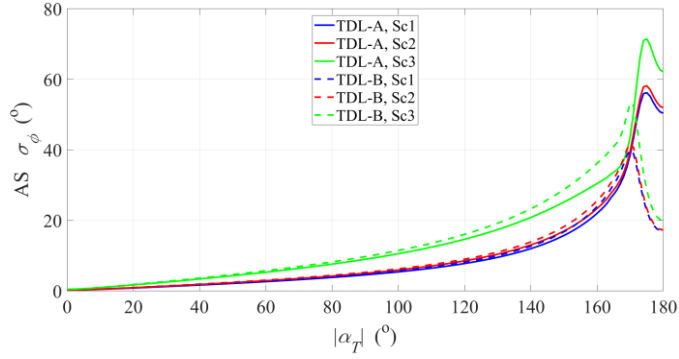

Fig. 2. AS versus $|\alpha_T|$ for indoor office and NBA.

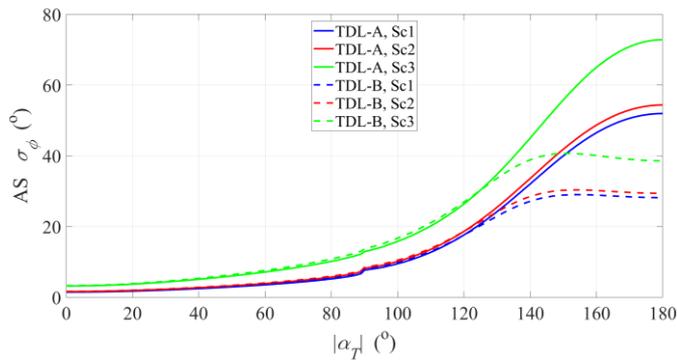

Fig. 3. AS versus $|\alpha_T|$ for indoor office and WBA.

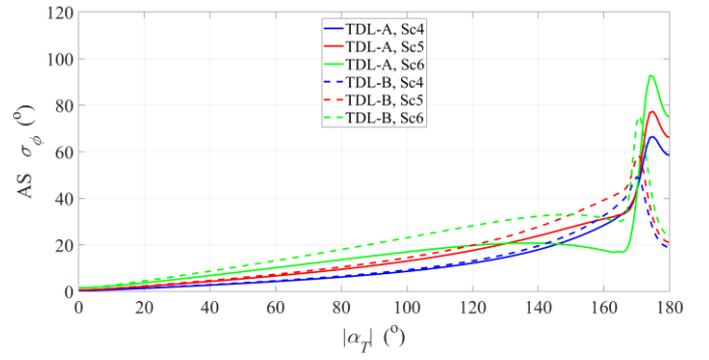

Fig. 4. AS versus $|\alpha_T|$ for UMi street-canyon and NBA.

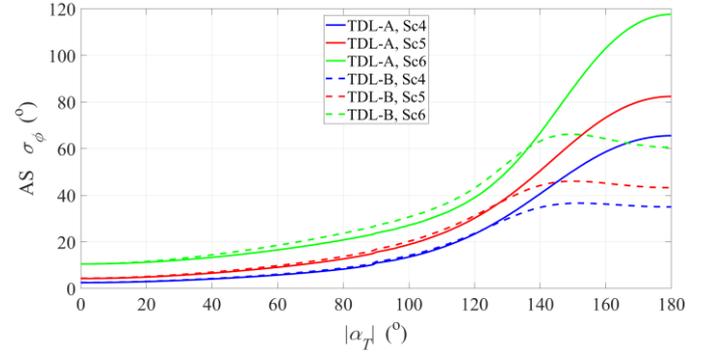

Fig. 5. AS versus $|\alpha_T|$ for UMi street-canyon and WBA.

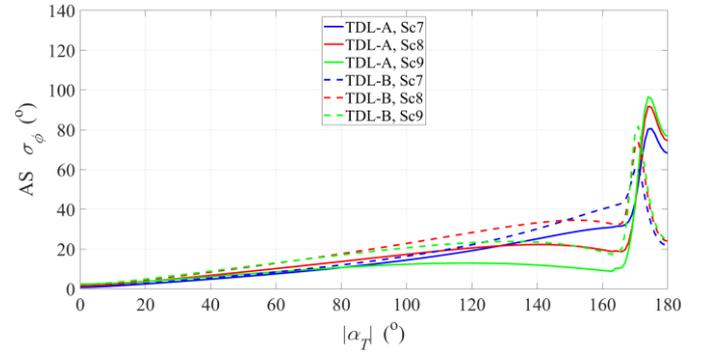

Fig. 6. AS versus $|\alpha_T|$ for UMa and NBA.

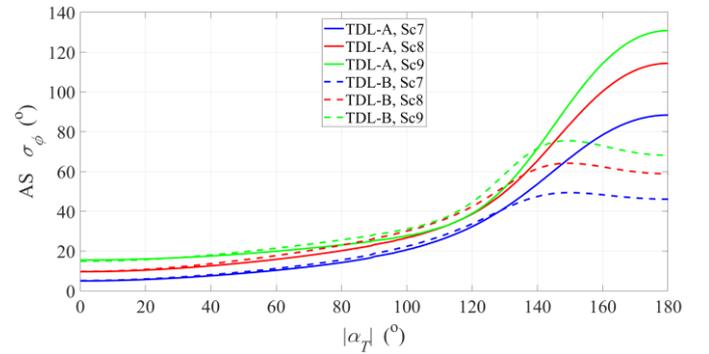

Fig. 7. AS versus $|\alpha_T|$ for UMa and WBA.

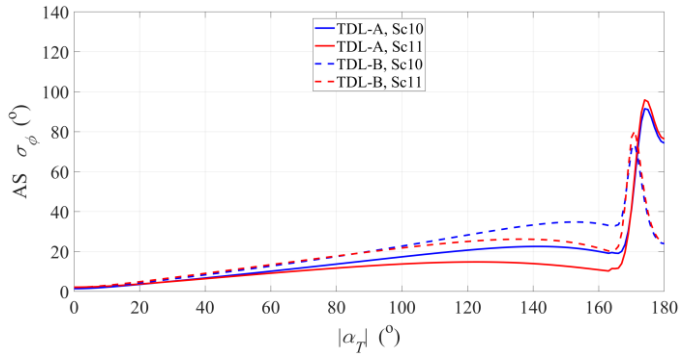

Fig. 8. AS versus $|\alpha_T|$ for UMi/UMa O2I and narrow-beam Tx antenna.

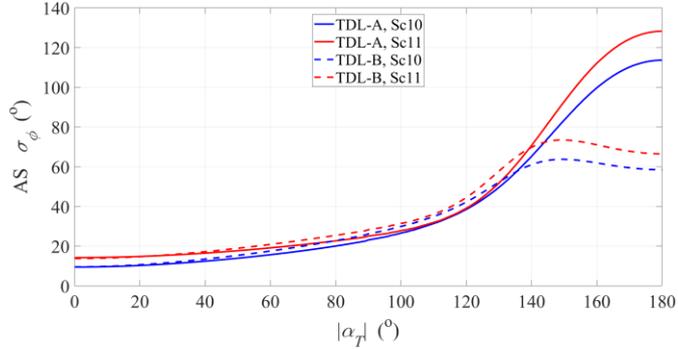

Fig. 9. AS versus $|\alpha_T|$ for UMi/UMa O2I and wide-beam Tx antenna.

We can see that the shape of the depicted graphs is closely related to the transmitting antenna parameters, especially its HPBW. Whereas, for fixed $\alpha_T$, an influence of PDP and DS on ASs is better visible. To compare the analyzed propagation environment types, ASs for $\alpha_T = 180°$ are presented in Table II.

TABLE II. COMPARISON OF AS AT RECEPTION POINT FOR DIFFERENT PROPAGATION SCENARIOS

| Environment Type | Scenario | DS (ns) | AS at reception point (°) | | | |
|---|---|---|---|---|---|---|
| | | | PDP: TDL-A | | PDP: TDL-B | |
| | | | NBA | WBA | NBA | WBA |
| Indoor office | Sc1 | 16 | 50.4 | 52.0 | 16.9 | 28.2 |
| | Sc2 | 18 | 52.0 | 54.4 | 17.3 | 29.4 |
| | Sc3 | 41 | 62.1 | 72.8 | 20.0 | 38.6 |
| UMi street-canyon | Sc4 | 30 | 58.5 | 65.6 | 19.0 | 35.0 |
| | Sc5 | 61 | 66.1 | 82.4 | 21.1 | 43.2 |
| | Sc6 | 297 | 74.9 | 117.6 | 24.2 | 60.5 |
| UMa | Sc7 | 78 | 68.2 | 88.4 | 21.8 | 46.1 |
| | Sc8 | 249 | 74.4 | 114.4 | 24.0 | 58.9 |
| | Sc9 | 786 | 76.6 | 130.8 | 24.9 | 68.1 |
| UMi / UMa O2I | Sc10 | 240 | 74.3 | 113.7 | 23.9 | 58.5 |
| | Sc11 | 616 | 76.3 | 128.3 | 24.8 | 66.5 |

In general, AS increase with increasing DS. The rate of increase depends on the used PDP and HPBW of the Tx antenna. Obviously, ASs are larger for WBA than NBA. In addition, ASs are greater for TDL-A than TDL-B. This is due to the fact that the power level for the local scattering, i.e., for delay equal to 0, in TDL-B (–13.4 dB) is lower than for TDL-A (0 dB).

For large DS, the AS increase is insignificant and we can assume that goes asymptotically to a certain limit value. This trend is well illustrated in Fig. 10.

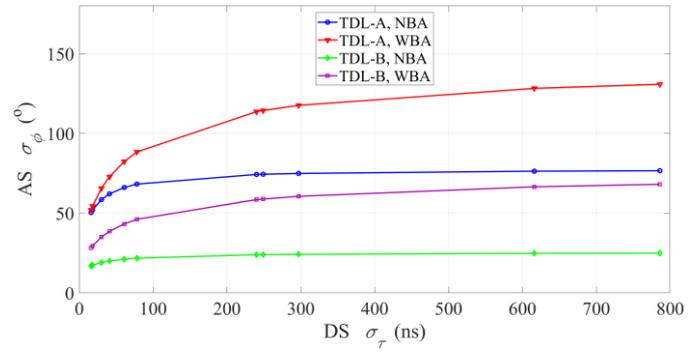

Fig. 10. AS versus DS for $\alpha_T = 180°$.

In [27], for 6 and 60 GHz, the influence of the directions, $\alpha_T$ and $\alpha_R$, of the transmitting and receiving antenna, respectively, on ASs at the reception point and receiving antenna output is analyzed. The obtained results show a significant influence of the directional receiving antennas on the reduction of the angular dispersion in the received signal. This fact is confirmed by the results presented in Table III, which are obtained for the analyzed environments and selected antenna directions, i.e., for $\alpha_T = 180°$ and $\alpha_R = 0°$.

TABLE III. COMPARISON OF AS AT RECEIVING ANTENNA OUTPUT FOR DIFFERENT PROPAGATION SCENARIOS

| Environment Type | Scenario | DS (ns) | AS at Rx antenna output (°) | | | |
|---|---|---|---|---|---|---|
| | | | TDL-A | | TDL-B | |
| | | | NBA | WBA | NBA | WBA |
| Indoor office | Sc1 | 16 | 3.5 | 15.4 | 3.5 | 9.1 |
| | Sc2 | 18 | 3.5 | 15.9 | 3.5 | 9.2 |
| | Sc3 | 41 | 3.5 | 19.5 | 3.5 | 9.6 |
| UMi street-canyon | Sc4 | 30 | 3.5 | 18.3 | 3.5 | 9.5 |
| | Sc5 | 61 | 3.5 | 20.5 | 3.5 | 9.6 |
| | Sc6 | 297 | 3.5 | 11.1 | 3.5 | 8.1 |
| UMa | Sc7 | 78 | 3.5 | 20.4 | 3.5 | 9.5 |
| | Sc8 | 249 | 3.5 | 12.4 | 3.5 | 8.3 |
| | Sc9 | 786 | 3.5 | 7.6 | 3.5 | 7.3 |
| UMi / UMa O2I | Sc10 | 240 | 3.5 | 12.7 | 3.5 | 8.4 |
| | Sc11 | 616 | 3.5 | 7.9 | 3.5 | 7.5 |

The influence of the antenna pattern on limiting the angular dispersion is particularly visible to NBA. In this case, AS is fixed, amount to 3.5°, and does not depend on the propagation environment type. For WBA, AS at the Rx antenna output in relation to the reception point can be from a few to a dozen times smaller.

V. CONCLUSION

The purpose of this paper was to evaluate the angular dispersion for various types of propagation environments for 38-39 GHz. The carried out simulation analysis was based on MPM and the 3GPP standard, which defines the environment types. The obtained results show a significant differentiation of AS at the reception point for various environments and transmitting antenna types. In addition, the significant effect of the directional receiving antennas on the reduction of the angular dispersion is shown. Presented issues are important from the point of view of the emerging 5G systems, which will be used the millimeter waves and new antenna techniques such as massive MIMO or APAA.